\documentclass[prb,aps,epsf,twocolumn,showpacs,10pt]{revtex4-1}
\usepackage{graphicx,amsfonts,times,bm,amsmath,verbatim,color,array}
\usepackage{graphicx,subfigure,epsfig}
\usepackage{dcolumn}
\usepackage{amssymb}
\usepackage{times}
\usepackage{amsmath}
\usepackage{amsfonts}
\usepackage{mathrsfs}
\usepackage{setspace}
\usepackage{latexsym}
\usepackage{bbm}
\usepackage{float}
\usepackage{flafter}
\usepackage{bm}
\usepackage{epstopdf}
\usepackage{hyperref}
\usepackage{color}
\usepackage{multirow}
\usepackage[bottom]{footmisc}

\def \be {\begin{eqnarray}}
\def \ee {\end{eqnarray}}

\newcommand{\ket}[1]{| {#1} \rangle}

\usepackage{subfigure}
\usepackage{float}
\usepackage{epsfig}
\usepackage{hyperref}
\hypersetup{
colorlinks=true,final=true,
        linkcolor=blue,
        citecolor=blue,
        filecolor=blue,
        urlcolor=blue,
}
\begin{document}
\title{Symmetry and Quantum Kinetics of the Non-linear Hall Effect}
%\title{Non-linear Hall effect Beyond the Berry curvature Dipole}

\author{S. Nandy$^{1,2}$}
\author{Inti Sodemann$^{1}$}
\affiliation{
$^{1}$ Max-Planck Institute for the Physics of Complex Systems, D-01187 Dresden, Germany\\
$^2$ Department of Physics, Indian Institute of Technology Kharagpur, W.B. 721302, India}

\begin{abstract}

We argue that the static non-linear Hall conductivity can always be represented as a vector in two-dimensions and as a pseudo-tensor in three-dimensions independent of its microscopic origin. In a single band model with a constant relaxation rate this vector or tensor is proportional to the Berry curvature dipole~\cite{Sodemann_2015}. Here, we develop a quantum Boltzmann formalism to second order in electric fields. We find that in addition to the Berry Curvature Dipole term, there exist additional disorder mediated corrections to the non-linear Hall tensor that have the same scaling in impurity scattering rate. These can be thought of as the non-linear counterparts to the side-jump and skew-scattering corrections to the Hall conductivity in the linear regime. We illustrate our formalism by computing the different contributions to the non-linear Hall conductivity of two-dimensional tilted Dirac fermions.
\end{abstract}


\maketitle

{\color{blue}{\em Introduction}}---Two independent experimental studies have recently reported the discovery~\cite{Mak2018,Pablo2018} of the time-reversal-invariant non-linear Hall effect (NLHE) in layered transition metal dichalcogenides. Unlike the ordinary Hall effect, the NLHE can occur in time-reversal-invariant metals lacking inversion symmetry~\cite{Spivak2009,MooreOrenstein2010,Sodemann_2015,Low2015}. Building upon previous studies~\cite{Spivak2009,MooreOrenstein2010} a simple semiclassical theory of this effect was developed in Ref.~\cite{Sodemann_2015} based on the notion of the {\it Berry curvature dipole} (BCD): a tensorial object measuring the average gradient of the Berry curvature over the occupied states. In a single band model with a constant relaxation rate the non-linear conductivity of a time reversal invariant metal was found to be proportional to the BCD. Several subsequent studies have addressed
the NLHE and related effects in a variety of contexts and material platforms~\cite{Nagaosa_2016, Xu_2018, Zhang_2018, Tsirkin_2018, Brink_2018, You_2018, Shi_2019, Facio_2018, Yan_2018, Lu_2018, Ryota_2018}. 

In this paper we elaborate further on the theory of the NLHE. We argue that the non-linear Hall conductivities can always be described by a pseudo-tensor in three-dimensions and a pseudo-vector in two-dimensions independent of their microscopic origin. This implies, that constraints imposed by crystal symmetry on the non-linear Hall pseudo-tensors/vectors follow from generic principles and in particular they are identical to the constraints imposed by crystal symmetry on the BCD pseudo-tensor/vector identified in Ref.~\cite{Sodemann_2015}. We study the NLHE within the quantum Boltzmann approach developed in Ref.~\cite{MacDonald_2017} which is able to capture a variety of phenomena in multi-band systems~\cite{Culcer_2017}. We have encountered several contributions to the NLHE beyond the BCD contribution. Notably, we have found non-linear side-jump (NLSJ) and skew-scattering (NLSK) terms, analogous to the corrections of the linear Hall conductivity~\cite{Nagaosa2010}, which have the same scaling with the scattering rate as the BCD contribution. Our approach also captures the intrinsic NLHE allowed in time-reversal broken metals identified in Ref.~\cite{Niu_2014}.

{\color{blue}{\em Symmetry constraints on the non-linear Hall effect}}---Let us consider a metal in a steady state flow of electric current in the presence of DC electric fields. To second order in electric fields the linear and non-linear conductivity tensors capture the electric current response:

%=======================================================================================================================================================================
\begin{figure}[t]
\begin{center}
\epsfig{file=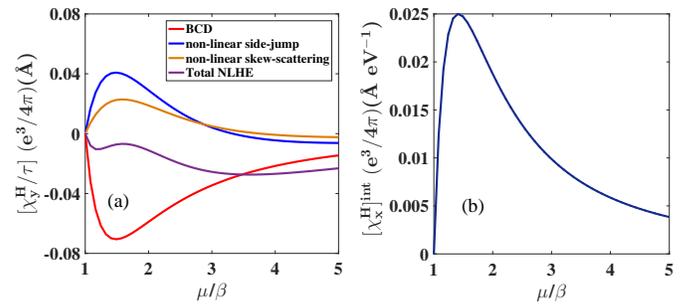,trim=0.0in 0.05in 0.0in 0.05in,clip=true, width=90mm}\vspace{0em}
\caption{(Color online) (a) Linear in $\tau$ contributions  to the non-linear Hall conductivity: Berry curvature dipole (BCD), non-linear side-jump (NLSJ), and non-linear skew-scattering (NLSK) as a function of chemical potential for a tilted Dirac cone. (b) Intrinsic contribution of the NLHC. Here, we have taken $\hbar=1$, $v=1$ eV$\cdot{\mbox{\normalfont\AA}}$, $\beta=0.6$ eV and $\alpha=0.3$ eV$\cdot{\mbox{\normalfont\AA}}$.}
%(units $\frac{e^3 \alpha}{4\pi\beta^2}$), (units $\frac{e^3\tau \alpha }{4\pi \beta}$)}
\label{NLH}
\end{center}
\end{figure}
%%======================================================================================
\be
j_{\alpha}=\sigma_{\alpha\beta} E_\beta+\chi_{\alpha\beta\gamma} E_\beta E_\gamma+\cdots
\ee

%\be
%j_{\alpha}=\sigma_{\alpha\beta} E_\beta+\sigma_{\alpha\beta\gamma} E_\beta E_\gamma+\cdots
%\ee

\noindent Here $\alpha$ refers to space indices and sum over repeated indices is understood. The power supplied by the electric field on the electronic fluid is the scalar $p=j_{\alpha} E_\alpha$. We wish to separate the conductivity tensors into the components that contribute to the power and dissipationless or Hall components. All the transformation properties discussed in this section follow from the elementary fact that the current and electric fields transform as vectors under coordinate changes. Before considering the non-linear response, we briefly review the case of the linear conductivity $\sigma_{\alpha\beta}$, where the decomposition amounts to separating the tensor into symmetric and antisymmetric components. In 2D there is a single independent component characterising the antisymmetric conductivity, which is the Hall conductivity, and transforms as a pseudo-scalar under a spatial symmetry coordinate changes:

\be\label{2DL}
{\rm2D:\ }\sigma^H \equiv \frac{\varepsilon_{\alpha\beta} \sigma_{\alpha\beta}}{2}, \ \sigma^H={\rm det} (O) \sigma^H,
\ee

\noindent here $\varepsilon_{\alpha\beta}$ is the 2D Levi-Civita symbol and $O$ is an orthogonal matrix describing a symmetry transformation. Equation~\eqref{2DL} implies the well-known constraint that the Hall conductivity vanishes when the system has a left-handed symmetry (${\rm det} (O)=-1$) such as a mirror plane. In 3D there are three independent components of the Hall conductivity that transform as a pseudo-vector:

\be\label{3DL}
{\rm3D:\ }\sigma^{H}_{\gamma} \equiv \frac{\varepsilon_{\gamma\alpha\beta} \sigma_{\alpha\beta}}{2}, \ \sigma^{H}_{\alpha}={\rm det} (O) O_{\alpha\beta} \sigma^{H}_{\beta}.
\ee

\noindent Here $\varepsilon_{\alpha\beta\gamma}$ is the 3D Levi-Civita symbol. Equation~\eqref{3DL} leads to useful constraints, such as that the Hall vector must be normal to any mirror plane, and, that the presence of two independent mirror planes would force all its components to vanish.

Let us now consider the case of the non-linear Hall conductivity tensor $\chi_{\alpha\beta\gamma}$. In order to isolate the components that do not contribute to dissipation it suffices to antisymmetrise the first index with either the second or third. These two choices of antisymmetrization are equivalent because, by construction, this tensor is symmetric under the last two indices $\chi_{\alpha\beta\gamma}=\chi_{\alpha\gamma\beta}$. Thus,  one obtains that in 2D  the second order Hall response has two independent components that transform as a pseudo-vector~\cite{note1} under space symmetries:

\be\label{2D}
{\rm2D:\ }\chi^{H}_{\gamma} \equiv \frac{\varepsilon_{\alpha\beta} \chi_{\alpha\beta\gamma}}{2}, \ \chi^{H}_{\alpha}={\rm det} (O) O_{\alpha\beta} \chi^{H}_{\beta}.
\ee

\noindent We will refer to $\chi^{H}_{\gamma}$ as the non-linear Hall vector. This transformation rule implies that the non-linear Hall vector in 2D is always orthogonal to the mirror planes, and that two or more mirrors would force it to identically vanish. In 3D one finds that there are 9 independent components of the non-linear Hall conductivity that transform as a rank 2 pseudotensor:

\be\label{3D}
{\rm3D:\ }\chi^{H}_{\gamma\eta} \equiv \frac{\varepsilon_{\alpha\beta\gamma} \chi_{\alpha\beta\eta}}{2}, \\ \chi^{H}_{\alpha\beta}={\rm det} (O) O_{\alpha\alpha'} O_{\beta\beta'} \chi^{H}_{\alpha'\beta'}.
\ee

\noindent We will refer to $\chi^{H}_{\gamma\eta}$ as the non-linear Hall tensor. The constraints imposed by different space groups on the non-linear Hall vector and tensor follow from these transformation laws and are identical to those of the BCD described in Ref.~\cite{Sodemann_2015}. In fact, in the case of the single band model with a constant relaxation rate of Ref.~\cite{Sodemann_2015}, there is a simple relation between the non-linear Hall and the BCD tensors, which, after symmetrising the DC conductivities from Ref.~\cite{Sodemann_2015}, reads as:

\be
{\rm2D:\ } \chi^{H}_{\alpha}=\frac{3e^3 \tau }{4} D_{\alpha},\\
{\rm3D:\ } \chi^{H}_{\alpha\beta}=\frac{3e^3 \tau }{4} \left(D_{\beta\alpha}-\frac{1}{3}  \text{Tr}(D) \delta _{\alpha\beta }\right).
\ee 

\noindent Thus the non-linear Hall vector is proportional to the BCD vector in 2D, while in 3D it is proportional to the traceless part of the BCD pseudo-tensor. As we will see, there are further terms contributing to the non-linear Hall tensor beyond the BCD. Importantly, some of them can also be linear in the scattering rate $\tau$.

{\color{blue}{\em Quantum Kinetic Framework}}---We consider electrons moving in a periodic crystal with Bloch states $|u^n_k\rangle$ and energy $\epsilon_k^n$, where $k$ is the crystal momentum and $n$ the band index. Electrons also experience a static disorder potential, $U(r)$, and an external constant electric field, $ E$. As described in Ref.~\cite{MacDonald_2017}, the disorder averaged density matrix or equal-time Green's function, satisfies a quantum Boltzmann equation of the form:
\begin{eqnarray}\label{c01}
\frac{d\rho}{dt}+\frac{i}{\hbar}[H_{0},\rho]+I(\rho)=-\frac{i}{\hbar}[H_{E},\rho],
\end{eqnarray}
where $\rho=\langle c^\dagger_{nk} c_{n'k}\rangle$. We employ the gauge in which the electric field is coupled linearly to the position operator, and we use the representation of this operator in the Bloch basis~\cite{Blount_1962, Aversa_1995}:
\begin{eqnarray}
r_{m m^{\prime}}=i\partial_k+A_{m m^{\prime}}(k),
\label{c02}
\end{eqnarray}
where $A_{m m^{\prime}}(k)=i\langle u_{k}^m|\partial_k|u_{k}^{m^{\prime}}\rangle$ is the non-Abelian Berry connection matrix. The collision operator is given by :
\begin{align}
&I(\rho)^{mm^{\prime \prime \prime}}_{\mathbf{k}}=\frac{\pi n_i}{\hbar} \sum_{m^{\prime} m^{\prime \prime}\mathbf{k}^{\prime}}\{ U_{\mathbf{k}\mathbf{k}^{\prime}}^{mm^{\prime}} U_{\mathbf{k}^{\prime}\mathbf{k}}^{m^{\prime}m^{\prime \prime}} \delta(\epsilon_{\mathbf{k^\prime}}^{m^\prime}-\epsilon^{m^{\prime \prime}}_{\mathbf{k}}) \rho^{m^{\prime \prime}m^{\prime \prime \prime}}_{\mathbf{k}} \nonumber \\
&-U_{\mathbf{k}\mathbf{k}^{\prime}}^{mm^{\prime}} U_{\mathbf{k}^{\prime}\mathbf{k}}^{m^{\prime \prime}m^{\prime \prime \prime}} \rho^{m^{\prime}m^{\prime \prime}}_{\mathbf{k^\prime}} \delta(\epsilon_{\mathbf{k^\prime}}^{m^{\prime \prime}}-\epsilon^{m^{\prime \prime \prime}}_{\mathbf{k}})-U_{\mathbf{k}\mathbf{k}^{\prime}}^{mm^{\prime}} U_{\mathbf{k}^{\prime}\mathbf{k}}^{m^{\prime \prime}m^{\prime \prime \prime}} \nonumber \\
& \delta(\epsilon_{\mathbf{k}}^{m}-\epsilon^{m^{\prime}}_{\mathbf{k^\prime}})\rho^{m^{\prime}m^{\prime \prime}}_{\mathbf{k^\prime}}+U_{\mathbf{k}\mathbf{k}^{\prime}}^{m^\prime m^{\prime \prime}} U_{\mathbf{k}^{\prime}\mathbf{k}}^{m^{\prime \prime}m^{\prime \prime \prime}} \rho^{m m^{\prime}}_{\mathbf{k}} \delta(\epsilon_{\mathbf{k}}^{m^\prime}-\epsilon^{m^{\prime \prime}}_{\mathbf{k^\prime}})\}. \nonumber \\
\label{e2}
\end{align}
Here $n_i$ is the impurity density and $U_{\mathbf{k}\mathbf{k}^{\prime}}^{mm^{\prime}}=\langle \psi_{{\mathbf{k}}^\prime}^{m^\prime}| U(r) | \psi_{\mathbf{k}}^m \rangle$, where $\psi_{\mathbf{k}}^m(r) = e^{ikr} u_{\mathbf{k}}^m(r)$. Notice that the collision operator is a linear functional of the density matrix and satisfies the fundamental property that it vanishes when evaluated in any diagonal density matrix that is a function of the energy $\rho_{\mathbf{k}}^{m m^\prime}=\delta_{m m^\prime} \rho_{0}(\epsilon_{\mathbf{k}}^m)$. This guarantees that the equilibrium Fermi-Dirac distribution is a solution of the quantum Boltzmann equation (QBE) in the absence of external electric field.

We solve the QBE perturbatively by performing a double expansion on the impurity density $n_{i}$, which controls the disorder strength, and the driving electric field $E$. To do so, we write the density
matrix as $\rho = \sum_{q,p}\rho_{q,p},$ where $\rho_{q,p}$ is understood to vanish as $ E^q n_i^p$. The series must start at $q=0$, namely $\rho_{q<0,p} = 0$, for any $p$. Moreover the $E^0$ term must coincide with the equilibrium distribution which is independent of the impurity strength. Therefore $\rho_{0,0} = \rho_{0}$ and $\rho_{0,p\neq 0} = 0$. We expect that for any $q$ , $\rho_{q,p}$ will have a strongest singularity of the form $1/n^q$ . This singularity encodes the fact in the clean limit the distribution is unable to reach a steady state in the absence of collisions. The largest power of this divergence follows from the expectation that the distribution has a leading correction at order $q$ of the form $E^q \tau^q$, where $\tau \sim 1/n_i$ is the relaxation time. Therefore, the expansion takes the form:
\begin{eqnarray}
\rho=\rho_{0}+\sum_{q \geq 1}\sum_{p \geq -q} \rho_{q,p}.
\label{e3}
\end{eqnarray}
The QBE can be solved recursively. The detailed recursive solution is described in the supplementary
material~\cite{Supple}.
We present here expressions for density matrices to second order in electric field, $\{\rho_{2,-2},\rho_{2,-1},\rho_{2,0}\}$, in terms of the density matrices linear in electric field $\{\rho_{1,-1},\rho_{1,0}\}$. Expressions for the latter are derived in Ref.~\cite{MacDonald_2017,Supple}. Our results can be used for any band structure with quenched disorder. The leading term in scattering rate is a diagonal matrix and given by:
\begin{eqnarray}\label{R14}
\rho_{2,-2}={\rho_{D}}_{2,-2}=-\frac{i}{\hbar}I_{D}^{-1}([e\mathbf{E}\cdot \hat{\mathbf{r}},\rho_{1,-1}]_{D}),
\end{eqnarray}
here $I_D^{-1}$ is the inverse of the collision operator viewed as an ordinary matrix acting on the diagonal part of the density matrix viewed as a vector~\cite{MacDonald_2017}. This term is the leading semiclassical term from Boltzmann theory that scales as $\tau^2$~\cite{Sodemann_2015}. The next correction is ${\rho}_{2,-1}$ which scales as $\tau$, and contains several effects that are the focus of our study. First, its band off-diagonal part ($m\neq m'$), is the sum of two matrices that give rise to the BCD and NLSJ terms:

\begin{eqnarray}\label{rBCD}
({\rho}_{\rm BCD})^{mm^{\prime}}_{\mathbf{k}}&&=- \frac{[e\mathbf{E}\cdot \hat{\mathbf{r}},\rho_{1,-1}]_{\mathbf{k}}^{mm^{\prime}}}{\epsilon_{m\mathbf{k}}-\epsilon_{m^{\prime}\mathbf{k}}},
\end{eqnarray}
\begin{eqnarray}\label{rNLSJ}
(\rho_{\rm NLSJ})^{mm^{\prime}}_{\mathbf{k}}&&=i\hbar \frac{I(\rho_{2,-2})_{\mathbf{k}}^{mm^{\prime}}}{\epsilon_{m\mathbf{k}}-\epsilon_{m^{\prime}\mathbf{k}}},
\end{eqnarray}

\noindent where we assume no band degeneracies at the fermi surface. Second, its band diagonal part ($m = m'$), is the term giving rise to the NLSK:

\begin{eqnarray}\label{rNLSK}
{\rho}_{\rm NLSK}=-I_{D}^{-1}\left(\frac{i}{\hbar}[e\mathbf{E}\cdot \hat{\mathbf{r}},{\rho}_{1,0}]_{D}+I_{D}({\rho_{OD}}_{2,-1})\right).
\label{R17}
\end{eqnarray}

\noindent where ${\rho_{OD}}_{2,-1}={\rho}_{\rm BCD}+\rho_{\rm NLSJ}$. Equations~\eqref{rBCD}-\eqref{rNLSK} contain all contributions up to order $\tau$. One can continue the expansion with the sub-leading matrix ${\rho}_{2,0}$ and several terms independent of $\tau$ would appear. Among these, one encounters an intrinsic band-off-diagonal term that depends only on band properties and not on collisions, given by:

\begin{eqnarray}\label{rINT}
({\rho}_{\rm INT})^{mm^{\prime}}_{\mathbf{k}}&&=- \frac{[e\mathbf{E}\cdot \hat{\mathbf{r}},\rho_{1,0}^{int}]_{\mathbf{k}}^{mm^{\prime}}}{\epsilon_{m\mathbf{k}}-\epsilon_{m^{\prime}\mathbf{k}}}.
\end{eqnarray}

\noindent As we will see, this term gives rise to the intrinsic NLHE in time-reversal broken metals first identified in Ref.~\cite{Niu_2014}. Given the density matrix one can compute any observable. In particular, the electric current will be given by the average of the velocity, and thus the non-linear conductivity will be:  

\begin{eqnarray}
\chi_{\alpha \beta \gamma}=-e\frac{\partial^2 {\rm tr}[v_{\alpha} \rho]}{\partial E_\beta\partial E_\gamma}\lvert_{E_\alpha\rightarrow0}.
\label{TR}
\end{eqnarray}

%\begin{eqnarray}
%\chi_{\alpha \beta \gamma}=\frac{\partial^2 {\rm tr}[-ev_{\alpha}\langle\rho_{E_\beta E_\gamma}\rangle_{\mathbf{k}}]}{\partial E_\alpha\partial E_\gamma}\frac{{\rm Tr}[-ev_{\alpha}\langle\rho_{E_\beta E_\gamma}\rangle_{\mathbf{k}}]}{E_\beta E_\gamma} .
%\label{TR}
%\end{eqnarray}

{\color{blue}{\em Non-linear Hall conductivity of 2D Dirac Fermions}}---To apply our formalism we consider a model of tilted 2D Dirac cones, which captures the low energy properties of various Dirac materials, such as the surface of topological crystalline insulators~\cite{Fu_2012, Fu_2015} and strained transition metal dichalcogenides. Their effective Hamiltonian is~\cite{Sodemann_2015}:
\begin{eqnarray}
H(\mathbf{k})=vk_{x}\sigma_{y}-vk_{y}\sigma_{x}+\alpha k_{y}+\beta \sigma_{z},
\label{H_Dirac}
\end{eqnarray}
where $v$ is the Fermi velocity, $\beta$ the gap, and $\alpha$ the tilt. The dispersion is $\epsilon^{\pm}_k=\alpha k_y \pm \epsilon_{k}$, $\epsilon_{k}=\sqrt{v^{2}k^{2}+\beta^{2}}$. Using the following gauge for the Bloch states:
\begin{eqnarray}
\ket{u_{\mathbf{k}}}^{\pm}=\frac{1}{\sqrt{2}}\begin{pmatrix} \begin{array}{c} \sqrt{1\pm\frac{\beta}{\epsilon_{k}}} e^{-i\theta}\\ \pm i\sqrt{1\mp\frac{\beta}{\epsilon_{k}}} \end{array} \end{pmatrix},
\label{e12}
\end{eqnarray}
\noindent the non-abelian Berry connection vector is found to be:
\begin{eqnarray}
&[A_{\mathbf{k}}]^{x}=-\frac{\sin \theta}{2k}\sigma_{0}-\frac{\beta \sin \theta}{2\epsilon_{k}k}\sigma_{z}-\frac{v \sin \theta}{2\epsilon_{k}}\sigma_{x}-\frac{\beta v \cos \theta}{2\epsilon_{k}^{2}}\sigma_{y} \nonumber \\
&[A_{\mathbf{k}}]^{y}=\frac{\cos \theta}{2k}\sigma_{0}+\frac{\beta \cos \theta}{2\epsilon_{k}k}\sigma_{z}+\frac{v \cos \theta}{2\epsilon_{k}}\sigma_{x}-\frac{\beta v \sin \theta}{2\epsilon_{k}^{2}}\sigma_{y} \nonumber \\
\label{e13}
\end{eqnarray}
Assuming a simple model of disorder with a random distribution of delta function impurities of the form $U(\mathbf{r}) =\sum_{i} U_{0}\delta(\mathbf{r}-\mathbf{r}_{i})$, the scattering time $\tau$ is found to be $\frac{1}{\tau}=\frac{n_{i}U_{0}^{2}}{4\hbar}\frac{\mu}{v^{2}}\left(1+3\frac{\beta^2}{\mu^2}\right)$, where $\mu=\epsilon_{k_F}$ is the chemical potential taken to be in the conduction band. 

Before discussing the non-linear transport we briefly recapitulate the linear response regime. We take the tilt in this case to be zero $\alpha=0$, because the mass term $\beta$ alone is enough to produce the linear Hall effect. The intrinsic part of the off-diagonal density matrix and its associated Hall conductivity are found to be $(\mathbf{E}=E_x \hat{x})$~\cite{Supple}:
\begin{eqnarray}
({\rho_{OD}}_{1,0})^{int}_{\mathbf{k}}&&=-\frac{eE_{x}}{2\epsilon_{k}} \left[\frac{v \sin \theta}{2\epsilon_{k}}\sigma_{x}+\frac{\beta v \cos \theta}{2\epsilon_{k}^{2}}\sigma_{y}\right] \nonumber \\
&&\times \quad [f_{0}(\epsilon^{+}_{k})-f_{0}(\epsilon^{-}_{k})],
\label{e14}
\end{eqnarray}
\begin{eqnarray}
\sigma_{yx}^{int}=\frac{e^2}{4\pi} \frac{\beta}{\mu}.
\end{eqnarray}
The above is equivalent to the well-known result from the integral of the Berry curvature.
%Now, in order to calculate the extrinsic part of the off-diagonal density matrix, we need to calculate the diagonal density matrix $(\rho_{E,-1})_{\mathbf{k}}^{mm}$ first. 
%In our model,   i.e. $\delta(\epsilon_{k}^{-}-\mu)=0$ and therefore $(\rho_{E,-1})^{--}_{\mathbf{k}}=0$. 
The extrinsic part of the off-diagonal and diagonal density matrices are found to be~\cite{Supple}:
\begin{eqnarray}
({\rho_{OD}}_{1,0})^{ext}_{\mathbf{k}}&&=\frac{eE_{x}\epsilon_{k}^2}{\epsilon_{k}^2+3\beta^2}\left[3 \frac{\beta v^{3}k^{2} \cos \theta}{4 \epsilon_{k}^{4}}\sigma_{y}+\frac{k^{2}v^{3}\sin \theta}{4\epsilon^{3}_{k}} \sigma_{x}\right] \nonumber \\
&&\times \delta(\epsilon_k-\mu) 
\label{e15}
\end{eqnarray}
%Similarly, the diagonal density matrix ${\rho_{D}}_{1,0}$ is~\cite{Supple}:
\begin{eqnarray}
({\rho_{D}}_{1,0})^{ext}_{\mathbf{k}}=-\frac{eE_x}{2\epsilon_k} \frac{\beta v^2 k\sin \theta (\epsilon_k^2-\beta^2)}{(\epsilon_k^2+3\beta^2)^2} \delta(\epsilon_{k}-\mu)\sigma_z \nonumber \\
\label{N2}
\end{eqnarray}
From these one obtains respectively the side-jump and skew-scattering contributions to the Hall conductivity~\cite{Supple}: 
\begin{eqnarray}
\quad \sigma_{yx}^{sj}=\frac{e^2}{2\pi}\frac{\beta}{\mu} \frac{(\mu^2-\beta^2)}{(\mu^2+3\beta^2)}, \quad \sigma_{yx}^{sk}=\frac{e^2}{2\pi}\frac{\beta}{2\mu} \frac{(\mu^2-\beta^2)^2}{(\mu^2+3\beta^2)^2}. \nonumber \\
\label{Lin_AHC}
\end{eqnarray}
We observe that the side-jump contribution arises from the band off-diagonal part of ${\rho}_{1,0}$ while the skew-scattering contribution arises from the diagonal part of ${\rho}_{1,0}$. Our skew-scattering term differs by a factor of 3 with respect to that obtained in a diagrammatic non-crossing approximation~\cite{Macdonald_2007,Ado_2015} although it has the same dependence on the mass and chemical potential. We note in passing that there is a debate on the full form of the linear Hall conductivity of massive Dirac fermions in the presence of disorder, as recent studies have advocated that diagrams beyond the non-crossing approximation have the same scaling in $\tau$~\cite{Ado_2015, Pesin_2018}. Our formalism can be viewed as a reasonable compromise between simpler phenomenological approaches and fully microscopic descriptions of disorder. We expect additional corrections to be present in such fully microscopic descriptions and hope that future studies will systematically study all leading disorder mediated corrections to the non-linear Hall conductivity.
%=====================================================================================================================================%=======================================================================================================================================================================
%\begin{figure}[t]
%\centering
%\begin{tabular}{cc}
%\includegraphics[width=44mm]{BCD_SJ_SK_NLHE.eps}
%\includegraphics[width=44mm]{INT_NLHC.eps}
%\end{tabular}
%\caption{(Color online) a) Magnitude of BCD, side-jump and skew-scattering contributions of non-linear Hall conductivity as a function of chemical potential for a tilted massive Dirac cone described in Eq.~(\ref{H_Dirac}) for $\alpha=0.1v$. b) depicts purely intrinsic contribution of the NLHC. Here we consider $\hbar=1$ and $\alpha/\beta=0.2$.}
%\label{NLH}
%\end{figure}
%%==================================================================================================

We now turn our attention to the NLHE. We begin by considering the contribution arising from the band off-diagonal density-matrix in Eq.~\eqref{rBCD} that gives rise to the BCD term. From this matrix we find the following BCD contribution to the non-linear Hall vector defined in Eq.~(\ref{2D}) to leading order in the tilt $\alpha$~\cite{Supple}:
\begin{eqnarray}
[\chi^{H}_{y}]^{\rm BCD}&&=-\frac{e^3 \tau(\mu)}{4\pi}\frac{\alpha\beta}{\mu^4 (\mu^2+3\beta^2)}(2\mu^4+\mu^2\beta^2-3\beta^4). \nonumber \\ 
&&=-\frac{e^3 \tau}{4\pi}\frac{3\alpha\beta}{2\mu^4}(\mu^2-\beta^2). 
\label{Non_Lin_BCD}
\end{eqnarray}
%\begin{eqnarray}
%[\chi^{H}_{y}]^{\rm BCD}=-\frac{e^3 \tau(\mu)}{4\pi}\frac{\alpha\beta}{\mu^4 (\mu^2+3\beta^2)}(2\mu^4+\mu^2\beta^2-3\beta^4). \nonumber \\
%\label{Non_Lin_BCD}
%\end{eqnarray}

\noindent The x-component vanishes $[\chi^{H}_{x}]^{BCD}=0$, and the second line follows after approximating the scattering rate to be energy independent and recovers the result from Ref.~\cite{Sodemann_2015}. We now consider the off-diagonal density matrix $\rho_{2,-1}$ arising from the collision operator, which is given in Eq.~(\ref{rNLSJ}), and which by analogy with the linear case can be identified as the NLSJ contribution.  The y-component is ($[\chi^{H}_{x}]^{\rm NLSJ}=0$):
\begin{eqnarray}
[\chi^{H}_{y}]^{\rm NLSJ}&&=\frac{e^3 \tau(\mu)}{4\pi}\frac{\alpha\beta (\mu^2-\beta^2)}{\mu^4 (\mu^2+3\beta^2)^3}(9\beta^6+12\mu^2\beta^4 \nonumber \\
&&+21\mu^4\beta^2-2\mu^6). \nonumber \\
\label{Non_Lin_AHC}
\end{eqnarray}

\noindent From the diagonal correction to the density matrix $\rho_{2,-1}$ given in Eq.~(\ref{rNLSK}), we obtain the NLSK contribution to the non-linear Hall vector, whose y-component reads as ($[\chi^{H}_{x}]^{\rm NLSK}=0$):
\begin{eqnarray}
[\chi^{H}_{y}]^{\rm NLSK}=-\frac{e^3 \tau(\mu)}{4\pi}\frac{\alpha\beta (\mu^2-\beta^2)}{\mu^2 (\mu^2+3\beta^2)^3}((\mu^{2}-7\beta^2)^2-52\beta^4) \nonumber \\
\label{Non_Lin_AHC_sk}
\end{eqnarray}
Notice that the BCD, NLSJ, and NLSK are all fermi surface effects and hence vanish in the limit of zero carrier density. The behavior of these different terms as a function of chemical potential is shown in Fig.~\ref{NLH}(a).

We have focused so far on the leading contributions to the non-linear Hall vector that have the same scaling as the BCD term. To illustrate the generality of our formalism, we will now consider a sub-leading density matrix that is independent of $\tau$ and which an intrinsic band structure effect, described in Eq.~(\ref{rINT}). This leads to a non-linear Hall vector with x-component given by ($[\chi^{H}_{y}]^{\rm INT}=0$):
\begin{eqnarray}
&&[\chi^{H}_{x}]^{\rm INT}=\frac{e^3}{4\pi}\frac{\alpha}{4\mu^4}(\mu^2-\beta^2).
\label{Non_Lin_Int}
\end{eqnarray}
This intrinsic contribution coincides with that previously identified in Ref.~\cite{Niu_2014} originating from the field induced positional shift of Bloch electrons. Unlike the terms that we discussed before, this contribution is only present in time reversal broken system, vanishing after adding pairs of Dirac cones related by time-reversal symmetry. It is also a fermi surface effect which vanishes as the chemical goes into the gap, unlike the intrinsic linear Hall effect. This contribution will be sub-leading in clean systems ($\tau \rightarrow \infty$). It is notable that this intrinsic contribution has a non-linear Hall vector orthogonal to the tilt of the Dirac cone, in contrast to the BCD, NLSJ, and NLSK contributions. Its dependence on the chemical potential is shown in Fig.~\ref{NLH}(b).

{\color{blue}{\em Summary}}---We have argued that the static non-linear Hall conductivity can always be represented as a vector in two-dimensions and as a pseudo-tensor in three-dimensions. Therefore the effect has a characteristic angular dependence which is dictated by crystal symmetry and is independent of the microscopic origin of the non-linear Hall effect. Within a quantum Boltzmann formalism we have shown that in addition to the Berry curvature dipole term identified in Ref.~\cite{Sodemann_2015}, there are two generic additional disorder mediated contributions that analogous to the side-jump and skew-scattering terms in the linear case that we termed the non-linear side-jump (NLSJ) and non-linear skew-scattering (NLSK) contributions, which also scale linearly with the impurity scattering time. This highlights the importance of taking into account these terms in recent experiments~\cite{Mak2018,Pablo2018}. We also recovered the subleading intrinsic non-linear Hall effect allowed in time-reversal broken systems that was identified in Ref.~\cite{Niu_2014}. 

During the completion of our work other manuscripts discussing disorder effects in the non-linear Hall effect with some overlap with our results appeared~\cite{Pesin_2018, Du_2018, Fu_2018, Niu_2019}.

\textit{Acknowledgments:} We are thankful to Allan H. MacDonald, Liang Fu, Suyang Xu, Qiong Ma, Oles Matshyshyn, Jhih-Shih You, Elio Konig, Jeroen van den Brink, Jorge Facio and Jairo Sinova for valuable discussions. SN acknowledges Arghya Taraphder and Sumanta Tewari for useful discussions.

%\begin{thebibliography}{10}
%\bibitem{Supplement} See supplementary information.
%\end{thebibliography}
%\pagebreak
%
%\begin{widetext}
%
%\appendix
%\onecolumngrid
%\appendix
%
%\section{\large Supplementary Information}

%\end{widetext}

%%%%%%%%%%%%%%%%%%%%%%%%%%%%%%%%%%%%%%%%%%%%%%%%%%%%%%%%%%%%%%%%%%%%%%%%%
%
%\begin{comment}
%\pagebreak
%\begin{widetext}
%\appendix
%\onecolumngrid
%\appendix
%\section{\large Supplementary Information}

%\end{widetext}
%\end{comment}

\begin{thebibliography}{10}
\bibitem{note1} There is no fundamental difference between pseudo-vectors and vectors in 2D because from any vector $v_\alpha$ a pseudovector can be obtained as $\tilde{v}_\alpha=\varepsilon_{\alpha\beta} v_\beta$ and vice versa.
\end{thebibliography}
\end{document}